\documentstyle[12pt,prd,aps]{revtex}
\begin{document}

\draft

\title{Cosmic Necklaces and Ultrahigh Energy Cosmic Rays}
      
\author{ Veniamin  Berezinsky$^\dagger$,
         Alexander Vilenkin$^*$} 
         
\address{$^\dagger$INFN, Laboratori Nazionali del Gran Sasso, 67010 Assergi 
           (AQ) Italy} 
\address{ $^*$Institute of Cosmology, Department of Physics and
Astronomy
, Tufts University, Medford, MA 02155, USA} 

\maketitle

\begin{abstract}

Cosmic necklaces are hybrid topological defects consisting of
monopoles and strings, with two strings attached to each monopole.  We
argue that the cosmological evolution of necklaces may significantly
differ from that of cosmic strings.  The typical velocity of necklaces
can be much smaller than the speed of light, and the characteristic
scale of the network much smaller than the horizon.  We estimate the flux
of high-energy protons produced by monopole annihilation in
the decaying closed loops.  For some reasonable values of the
parameters it is comparable to the observed flux of ultrahigh-energy
cosmic rays.

\end{abstract}
\pacs{98.80.Cq, 98.70.Sa}
\vskip .5cm

The observation of cosmic ray particles with energies higher than $10^{11}~GeV$
\cite{EHE}  gives a serious challenge to the known mechanisms of acceleration. 
The shock acceleration in different 
astrophysical objects typically gives maximal energy of accelerated protons
less than $(1-3)\cdot 10^{10}~GeV$ \cite{NMA}. The unipolar induction can 
provide the maximal energy $1\cdot 10^{11}~GeV$ only for the extreme values 
of the parameters \cite{BBDGP}. Much attention has recently been given to 
acceleration by ultrarelativistic shocks \cite{Vie},\cite{Wax}. The
particles here can gain a tremendous increase in energy,
equal to $\Gamma^2$, at a single reflection, 
where $\Gamma$ is the Lorentz factor of the shock.
However, it is 
known (see e.g.  the simulation 
for pulsar relativistic wind in \cite{Hosh}) that particles entering 
the shock region are captured there or at least have a small probability 
to escape. 

{\em Topological defects} (for a review see \cite{Book}) can naturally 
produce particles of ultrahigh energies (UHE) well in excess of those 
observed in cosmic rays (CR). In most cases the problem with topological 
defects is not the maximal energy, but the fluxes.

{\em Cosmic strings} can produce particles when two segments of
string come 
into close contact, as in {\it cusp} events \cite{Bran}.  
When the distance between two segments of the cusp
becomes of the order of the string width, the cusp may``annihilate" 
turning into high energy particles.
However, the resulting cosmic ray flux is far too small
\cite{BBM}.  

{\em Superconducting strings} \cite{Witten} appear to
be much better suited for particle production.
Moving through cosmic magnetic fields, such strings develop
electric currents and copiously produce charged heavy particles when the
current reaches certain critical value. 
The CR flux produced by
superconducting strings is affected by some model-dependent string
parameters and by the history and spatial distribution of cosmic
magnetic fields.  
Models considered so far failed to account
for the observed flux \cite{HS,SJSB}.

{\em Monopole-antimonopole pairs } ($M{\bar M}$) 
can form bound states and eventually
annihilate into UHE particles  \cite{Hill}, \cite{BS}.  
For an appropriate choice of the
monopole density $n_M$, this model is consistent with observations;
however, the required (low) value of $n_M$ may be difficult to explain.

We shall consider here another potential source of UHE CR, 
the topological defects which have not been much studied so far: 
{\em cosmic necklaces}. Such defects can be formed 
in a sequence of symmetry breaking phase transitions 
$G\to H\times U(1)\to H\times Z_2$.  If the group $G$ is semisimple, 
then the first phase transition
produces monopoles, and at the second phase transition each monopole
gets attached to two strings, with its magnetic flux channeled along the 
strings.  The resulting necklaces resemble ``ordinary'' cosmic strings
with monopoles playing the role
of beads.  ``Realistic'' particle physics models with necklaces can
readily be constructed \cite{HK}. 

The evolution of necklaces is rather complicated, and its analysis
would require high-resolution numerical simulations.  Here we shall
attempt only to indicate the relevant physical processes and to give
very rough estimates for the efficiency of some of these processes.  

The monopole mass $m$ and the string tension $\mu$ are determined by
the corresponding symmetry breaking scales, $\eta_s$ and $\eta_m$ 
($\eta_m>\eta_s$), 
\begin{equation}
m\sim 4\pi\eta_m/e,\;\;\;\; \mu\sim 2\pi\eta_s^2.
\label{m}
\end{equation}
Here, $e$ is the gauge coupling.  The mass per unit length of string
is equal to its tension, $\mu$.  Each string attached to a monopole
pulls it with a force $F=\mu$ in the direction of the string.  
The monopole radius $\delta_m$ and the string thickness $\delta_s$ are
typically of the order $\delta_m\sim(e \eta_m)^{-1}$, $~~\delta_s\sim
(e \eta_s)^{-1}$.  

Monopoles are formed at a temperature $T_m\sim\eta_m$.  
Their initial average
separation, $d$, can range from $\delta_m$ (for a second-order
phase transition)  to the horizon size (for a
strongly first-order transition).  
The monopoles are diluted by the
expansion of the universe, so that $d$ grows as $d\propto T^{-1}$.
There is some additional decrease in the monopole density, and
associated increase in $d$, due to $M{\bar M}$ annihilation.  The
latter process, however, is rather inefficient.

At the second phase transition, each monopole gets attached to two
strings, resulting in the formation of necklaces.  There will be
infinite necklaces having the shape of random walks and a distribution
of closed loops.  The two strings attached to a monopole are pulling
it with an equal force; hence, there is no tendency for a monopole to
be captured by the nearest antimonopole , 
unless their separation is comparable to the string thickness, $\delta_s$.  

An important quantity for the necklace evolution  is the dimensionless ratio
\begin{equation}
r=m/\mu d,
\label{r}
\end{equation}
The average mass per unit length of
necklaces is $(r+1)\mu$.
The initial value of $r$ can be large ($r\gg 1$) or small ($r\ll 1$),
depending on the nature of the two phase transitions.  

We expect the necklaces to evolve in a 
scaling regime. If $\xi$ is the characteristic length scale of the network, 
equal to the typical separation of long strings and to their characteristic
curvature radius, then the force  per unit length of string is 
$f \sim \mu/\xi$, and the acceleration is $a \sim  
(r+1)^{-1}\xi^{-1}$.
We assume that $\xi$ changes on a Hubble time scale $\sim t$. Then the 
typical distance travelled by long strings in time $t$ should be 
$\sim \xi$, so that the strings have enough time to intercommute in a 
Hubble time. This gives $at^2 \sim\xi$, or 
\begin{equation}
\xi \sim (r+1)^{-1/2}t.
\label{xi}
\end{equation}
The typical string velocity is $v \sim (r+1)^{-1/2}$.

For $r\ll 1$ the monopoles are subdominant, and the string evolution
is essentially the same as that of `ordinary' strings without
monopoles.  The opposite case $r\gg 1$ is much
different: the string motion is
slow and their average separation is small.  Like ordinary strings, 
cosmic necklaces can serve as seeds for
structure formation.  Significant quantitative changes in the
corresponding scenario can be expected for $r\gg 1$.

Disregarding $M{\bar M}$ annihilation, 
the evolution of $r(t)$  can be analyzed using the energy
balance equation 
\begin{equation}
{\dot E}=-P{\dot V}-{\dot E}_g.
\label{energy}
\end{equation}
Here, $E$ is the energy of long necklaces in a co-moving volume $V$,
$P$ is the effective pressure, 
and ${\dot E}_g$ is the rate of energy loss by gravitational radiation
from small-scale wiggles on long strings.  
If the scale of the wiggles is set by the gravitational back-reaction,
then the strings radiate a substantial part of their energy in a
Hubble time \cite{Hindmarsh,AS}, and we can write ${\dot E}_g=\kappa_g
Nm/rt$ where $N$ is the number of monopoles in volume $V$ and 
$\kappa_g\sim 1$.  The effect of loop formation is not
relevant for the evolution of $r(t)$ and has not been included in
Eq.(\ref{energy}). 

For $r\ll 1$, the effect of monopoles on the string dynamics is
negligible, and we can write $P=(Nm/3Vr)(2v^2-1)$, where $v$ is the
{\it rms} string velocity.  Then, with a power-law expansion 
$a(t)\propto t^\nu$, we obtain the following equation for $r(t)$:
\begin{equation}
{{\dot r}\over{r}}=-{\kappa_s\over{t}}+{\kappa_g\over{t}},
\label{rr}
\end{equation}
where $\kappa_s=\nu(1-2v^2 )$.
The first term on the {\it rhs} 
of Eq.(\ref{rr}) describes the string stretching 
due to expansion of the Universe while the second term describes the 
competing effect of string shrinking due to gravitational radiation
\cite{Foot10}. 
In this regime, we can use the values of $v^2$ from the string
simulations \cite{BB}: $v^2=0.43$ in the radiation era and $v^2=0.37$
in the matter era.  The corresponding values of $\kappa_s$ are,
respectively, 0.07 and 0.14.  Our estimate for $\kappa_g$ is
$\kappa_g\sim 1$, so 
it seems reasonable to assume that $\kappa_g>\kappa_s$.  
The solution of Eq.(\ref{rr}) is $r(t) \propto t^{\kappa_g-\kappa_s}$, 
suggesting
that if $r$ is initially small, it will grow at least until it reaches
values $r\sim 1$.

An equation similar to (\ref{rr}) can also be written for $r\gtrsim 1$,
but in this case the results of numerical simulations \cite{BB} can no
longer be used, and the relative magnitude of $\kappa_s$ and
$\kappa_g$ cannot be assessed.  Order-of-magnitude estimates suggest
$\kappa_s\sim\kappa_g\sim 1$, and in this paper we shall assume that
$\kappa_g>\kappa_s$, so that $r(t)$ is driven towards large values,
$r\gg 1$.

As $r$ grows and monopoles get closer together, $M{\bar M}$
annihilation should become important at some point.  In any case, the
growth of $r$ should terminate at the value
$r_{max}\sim\mu/m\delta_s\sim\eta_m/\eta_s$, when the monopole
separation is comparable to the string thickness $\delta_s$.  It is
possible that annihilations will keep $r$ at a much smaller value.
For example, if monopoles develop appreciable relative velocities
along the string, they may frequently run into one another and
annihilate.  The terminal value of $r$ cannot be determined without
numerical simulations of network evolution; here we shall assume that
$r\gg 1$.

Self-intersections of long necklaces result in copious production of
closed loops.  For $r\gtrsim 1$ the motion of loops is not periodic,
so loop self-intersections should be frequent and their fragmentation
into smaller loops very efficient.
A loop of size $\ell$ typically disintegrates on a timescale 
$\tau\sim r^{-1/2}\ell$.  All monopoles trapped in the loop must, of course, 
annihilate in the end.

Annihilating $M{\bar M}$ pairs decay into
Higgs and gauge bosons, which we shall refer to collectively as
$X$-particles.  The rate of $X$-particle production is easy to
estimate if we note that infinite necklaces lose a substantial
fraction of their length to closed loops in a Hubble time.  
The string length per unit volume is $\sim \xi^{-2}$, and the monople
rest energy released per unit volume per unit time is
$r\mu/\xi^2 t$.  Hence, we can write
\begin{equation}
{dn_X\over{dt}}\sim {r^2\mu\over{t^3 m_X}},
\label{xrate}
\end{equation}
where $m_X\sim e\eta_m$ is the $X$-particle mass  and we have used
Eq.(\ref{xi}) .

In the extreme case of $r\sim r_{max} \sim \eta_m/\eta_s$,
Eq.(\ref{xrate}) 
gives the rate of X-particle production which does not depend on the string
scale $\eta_s$. It is possible that the 
evolution of $r(t)$ is actually saturated in this regime.

$X$-particles emitted by annihilating monopoles decay into hadrons,
photons and neutrinos, which contribute to the spectrum of cosmic ultra-high 
energy radiations. In particular the diffuse flux of ultra-high energy protons
can be evaluated as
\begin{equation}
I_p(E)={1\over{4\pi}}{dn_X\over{dt}}{\lambda_p(E)\over{m_X}}W_N(m_X,x),
\label{pflux}
\end{equation}
where $dn_X/dt$ is given by Eq.(\ref{xrate}), 
$\lambda_p(E)$ is the attenuation lengh for ultra-high energy 
protons due to their interaction with microwave photons and
$W_N(m_X,x)$ is the fragmentation function of X-particle into nucleons
of energy $E=xm_X$.

The fragmentation function is calculated using the decay of X-paricle into QCD 
partons (quark, gluons and their supersymmetric partners) with the 
consequent development of the parton cascade. The cascade in this case is
identical to one initiated by $e^+e^-$ -anihilation.
We have used  the fragmentation function in the gaussian form as 
obtained in MLLA approximation in \cite{DHMT} and
\cite{ESW}. Additionally, we took into 
account the supersymmetric corrections to the coupling constant $\alpha_s$
at large $Q^2$. The details will be described elsewhere. Here we shall 
give only the explicit form of the fragmentation function we used:
\begin{equation}
W_N(m_X,x)=\frac{K_N}{x}exp \left( -\frac{\ln^2 x/x_m}{2\sigma^2} \right),
\label{frag}
\end{equation}
where
$$
2\sigma^2=\frac{1}{6} \left( \ln \frac{m_X}{\Lambda} \right) ^{3/2},
$$
$x=E/m_X$, $x_m=(\Lambda/m_X)^{1/2}$, $\Lambda=0.234~GeV$ with the 
normalization constant $K_N$ to be found from energy conservation 
assuming that about $10\%$ of initial energy ($m_X$) is transferred to 
nucleons.


For attenuation length of UHE protons due to their interactions with
microwave photons we used the calculations described in the book 
\cite{BBDGP}. 

Note that in our calculations 
the UHE proton flux is fully determined by only two parameters,
$r^2\mu$ and $m_X$. 
The former is restricted by low energy diffuse gamma-radiation.
It results from e-m cascades initiated 
by high energy photons and electrons produced in 
the decays of X-particles.

The cascade energy density predicted in our model is 
\begin{equation}
\omega_{cas}=\frac{1}{2}f_{\pi}r^2\mu \int_0 ^{t_0}\frac{dt}{t^3}
\frac{1}{(1+z)^4}=\frac{3}{4}f_{\pi}r^2\frac{\mu}{t_0^2},
\label{cas}
\end{equation}
where $t_0$ is the age of the Universe (here and below we use
$h=0.75$), $z$ is the redshift and $f_{\pi} \sim 1$ is the fraction of
energy 
transferred to pions. In Eq.(\ref{cas}) we took into account that half 
of the energy of pions is transferred to photons and electrons. 
The observational bound on the cascade density, for the kind of
sources we are considering here, is \cite{B92} $\omega_{cas} \lesssim
10^{-5}~eV/cm^3$.  This gives a bound on the parameter $r^2\mu$.

In numerical  calculations  we used 
$r^2\mu= 1\times 10^{28}~GeV^2$, which results in 
$\omega_{cas}=5.6 \cdot 10^{-6}~eV/cm^3$, somewhat below the observational 
limit. Now we are left with one free parameter, $m_X$, which we fix at
$1\cdot 10^{14}~GeV$.  Note that with this value, the maximum energy
of protons is not very high: $E_{max} \sim 10^{13}~GeV$.
The calculated proton 
flux is presented in Fig.1, together with a summary 
of observational data taken 
from ref.\cite{akeno}. These data are usually interpreted as
indicating the presence of a new 
component at energy higher than $1\cdot 10^{10}~GeV$. One cannot claim
that our
predicted flux gives a good fit to the data for this component, but
the discrepancy does not exceed $2\sigma$. 

Let us now turn to the calculations of UHE gamma-ray flux from  the decays 
of X-particles. The dominant channel is given by the decays of neutral 
pions. The flux can be readily calculated as
\begin{equation}
I_{\gamma}(E)=\frac{1}{4\pi}\frac{dn_X}{dt}\lambda_{\gamma}(E)N_{\gamma}(E)
,
\label{gflux}
\end{equation}
where $dn_X/dt$ is given by Eq.(\ref{xrate}), 
$\lambda_{\gamma}(E)$ is the absorption length of a photon with energy 
$E$ due to $e^+e^-$ pair production on background radiation and 
$N_{\gamma}(E)$ is the number of photons with energy E produced per one decay 
of X-particle. The latter is given by 
\begin{equation}
N_{\gamma}(E)=\frac{2K_{\pi^0}}{m_X}\int_{E/m_X}^1 \frac{dx}{x^2}
\exp \left( -\frac{\ln^2 x/x_m}{2\sigma^2} \right)
\label{gnumber}
\end{equation}
The normalization constant $K_{\pi^0}$ is again found from the
condition that  
neutral pions take away $f_{\pi}/3$ fraction of the total energy $m_X$.

An important point in our calculations was accounting for the
absorption of UHE photons due to 
$e^+e^-$ production on background radiation. 
At energy $E>1\cdot 10^{10}~GeV$ the 
dominant
contribution to the absorption comes from the radio background. The 
significance of this process was first noticed in \cite{B70}(see also 
book \cite{BBDGP}). New calculations for this absorption 
were recently
done \cite{PB}. 
We have used the 
absorption lengths from this work. 

The calculated flux of gamma radiation is presented in Fig. 1 by the curve
labelled $\gamma$. One can see that at $E \sim 1\cdot 10^{11}~GeV$ the 
gamma ray flux is considerably lower than that of protons. This is 
mainly due 
to the difference in the attenuation lengths for protons ($110~Mpc$) and 
photons ($2.6~Mpc$ \cite{PB} and $2.2~Mpc$ \cite{B70}). At higher energy 
the attenuation length for protons dramatically decreases ($13.4~Mpc$ at 
$E=1 \cdot 10^{12}~GeV$) and the fluxes of protons and photons become 
comparable.

A requirement for the models explaining the observed UHE events is that 
the distance between sources must be smaller than the attenuation
length. Otherwise the flux at the 
corresponding energy would be 
exponentially suppressed. This imposes a severe constraint on the
possible sources.  For example, in the case of protons with energy 
$E \sim (2- 3)\cdot 10^{11}~GeV$ 
the proton 
attenuation length is $19~Mpc$ . If protons 
propagate rectilinearly, there should be several sources inside this
radius;
otherwise all particles would arrive from the same direction.
If particles are strongly deflected in extragalactic magnetic fields, 
the distance to the source should be even smaller.  Therefore, the 
sources of 
the observed events at the highest energy must be at a distance 
$R\lesssim 15~Mpc$ in the case or protons. 

In our model the distance between 
sources, given by Eq.(\ref{xi}), satisfies this condition for 
$r>3\cdot 10^{4}$,
while $r_{max} \sim \eta_m/\eta_s$ can be many orders of magnitude larger.
This is in contrast to other potential sources,
including supeconducting cosmic strings and powerful 
astronomical sources such as AGN, for which this condition imposes
severe restrictions.  

The difficulty is even more pronounced in the case of UHE photons. 
These particles 
propagate rectilinearly and their absorption length is shorter: 
$2 - 4~Mpc$ at $E \sim 3\cdot 10^{11}~GeV$. It is rather unrealistic to expect
several powerful astronomical sources at such short distances. This 
condition is very restrictive for topological defects as well. The
necklace model we introduced here is rather exceptional.
  
In conclusion, we do not claim that we found a model explaining 
the observations, but almost.

\begin{figure}
\caption{Predicted proton and gamma-ray fluxes from necklaces.  The
data points are fluxes from the compilation made in Ref.[23].}
\end{figure}


\begin{thebibliography}{99}
\bibitem{Foot10}
Friction due to interaction with
ambient plasma can also contribute to string shrinking.  However, we 
expect it to be subdominant at the epochs relevant for UHE CR 
production.
\end{thebibliography}
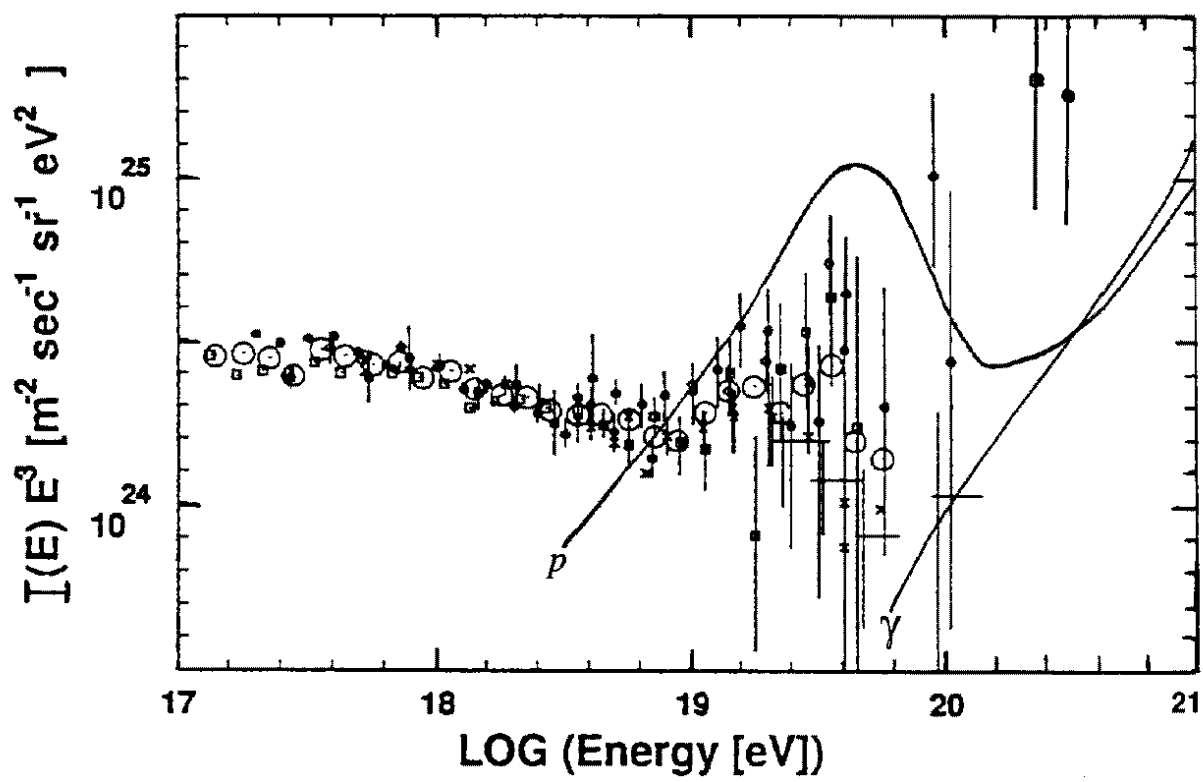
\end{document}